\documentclass[aps,prb,twocolumn,showpacs,floadfix,groupedaddress]{revtex4}
\usepackage{amsfonts}
\usepackage{graphicx}
\usepackage{color}
\usepackage{amsmath}
\usepackage{amssymb}
\usepackage{latexsym}

\begin{document}

\title{Effect of short- and long-range scattering in the conductivity of
graphene: Boltzmann approach vs tight-binding calculations}
\author{J. W. K{\l }os}
\email{klos@amu.edu.pl}
\affiliation{Surface Physics Division, Faculty of Physics, Adam Mickiewicz University,
Umultowska 85, 61-614 Pozna\~{n}, Poland}
\author{I. V. Zozoulenko}
\email{Igor.Zozoulenko@itn.liu.se}
\affiliation{Solid State Electronics, ITN, Link\"oping University 601 74, Norrk\"{o}ping,
Sweden }
\date{\today }

\begin{abstract}
We present a comparative study of the density dependence of the conductivity of graphene
sheets calculated in the tight-binding (TB) Landauer approach and on the basis of the
Boltzmann theory. The TB calculations are found to give the same density dependence of
the conductivity, $\sigma \sim n$, for short-range and long-range Gaussian scatterers. In
the case of short-range scattering the TB calculations are in agreement with the
predictions of the Boltzmann theory going beyond the Born approximation, but in
qualitative and quantitative disagreement with the standard Boltzmann approach within the
Born approximation, predicting $\sigma= $ const. Even for the long-range Gaussian
potential in a parameter range corresponding to realistic systems the standard Boltzmann
predictions are in quantitative and qualitative disagreement with the TB results. This
questions the applicability of the standard Boltzmann approach within the Born
approximation, commonly used for the interpretation of the results of experimental
studies of the transport in graphene.
\end{abstract}



\maketitle

\section{Introduction}

Understanding factors that affect conductivity of graphene represents a
fundamental task of great importance in view of possible application of
graphene-based devices for electronics and optoelectronics. Currently, the
majority of experimental measurements of conductivity $\sigma $ in graphene
\cite{Morozov,Tan,BolotinSSC08,BolotinPRL08,Avouris,Jang,Huard,Hong} is
analyzed on the basis the standard Boltzmann approach within the Born
approximation predicting qualitative different results for short- and
long-range impurity scattering\cite{Nomura,Ando2006,Hwang2007,Vasko,Stauber}
\begin{subequations}
\label{short-long}
\begin{eqnarray}
\sigma &=&\text{const (short-range scattering),}  \label{short} \\
\sigma &\sim &n\text{ (long-range scattering),}  \label{long}
\end{eqnarray}%
where $n$ is the electron density. It has been recently argued by Stauber \textit{et al}.
\cite{Stauber} that a standard way to examine the collision rate within the Born
approximation (utilizing the unperturbed wave functions for a clean system) is not
suitable for the case of short-range interaction such as vacancies, resonant impurities,
cracks etc. Going beyond the Born approximation, Stauber \textit{et al.}\cite{Stauber}
and Katsnelson and Novoselov\cite{Katsnelson} demonstrated that the short-range disorder,
with the accuracy up to logarithmic corrections, leads to a linear density dependency
similar to the one for the long-range potential,
\end{subequations}
\begin{equation}
\sigma =\frac{4e^{2}}{h}\frac{n}{n_{i}}\left( \ln \sqrt{\pi n}R_{0}\right)
^{2},  \label{Stauber}
\end{equation}%
where $R_{0}$ is the scatterer's radius, and $n_{i}$ is the impurity
concentration. (Similar results have been also obtained by Ostrovsky \textit{%
et al.}\cite{Ostrovsky}). Apparently, this has important consequence for interpretation
of the experimental results, as the linear density dependence of the conductivity is
typically related to the long-range Coulomb impurities, and deviations from this
dependence is attributed to the
short-range scattering.\cite%
{Morozov,Tan,BolotinSSC08,BolotinPRL08,Avouris,Jang,Huard,Hong} In contrast, Eq.
(\ref{Stauber}) implies the short- and long-range scattering may lead to similar density
dependencies of the conductivity. Indeed, Monteverde \textit{et al}.\cite{Monteverde}
have recently analyzed the experiment on the basis of Eq. (\ref{Stauber}) and arrived to
the conclusion that strong neutral defects (as opposed to the long-range Coulomb
impurities) was the main scattering mechanism in graphene. The dominant role
of neutral defects has been also recently outlined in Refs.%
\onlinecite{Ponomarenko,Ni}.

The reliability of the above predictions (\ref{short-long}), (\ref{Stauber})
can be established by testing them against \textquotedblleft
exact\textquotedblright\ Landauer-type quantum mechanical numerical
calculations for the conductivity based on the tight-binding (or Dirac)
Hamiltonian for carriers in graphene. \cite%
{Bardarson,Lewenkopf,Robinson,Xu,Xu_bilayer,ABDS,Klos}. Recently, Adam
\textit{et al.} \cite{ABDS} compared the standard Boltzmann and the Landauer
approaches for the case of a long-range Gaussian potential that varies
smoothly on the scale of a lattice constant,
\begin{equation}
V_{i}=\sum_{i^{\prime }=1}^{N_{imp}}U_{i^{\prime }}\exp \left( -\frac{|%
\mathbf{r}{_{i}}-\mathbf{r}_{i^{\prime }}|^{2}}{2\xi ^{2}}\right) ,
\label{Gauss_pot}
\end{equation}%
where $\xi $ can be interpreted as the effective screening length, and the
potential heights is assumed to be uniformly distributed in the range $%
U_{i}\in \lbrack -\delta ,\delta ]$. The conductivity obtained in the two approaches
agree quantitatively away from the Dirac point, which was interpreted as a proff of
validity of both. According to Adam \textit{et al.} \cite{ABDS}, the conductivity follows
a density dependence ~ $\sigma \sim n^{3/2}.$ This however disagrees with all
experimental
observations reported so far \cite%
{Morozov,Tan,BolotinSSC08,BolotinPRL08,Avouris,Jang,Huard,Hong,Monteverde,Ponomarenko,Ni,Du}
and with previous Landauer-type numerical calculations \cite%
{Robinson,Lewenkopf,ABDS,Klos} demonstrating the linear or sublinear density dependence
of the conductivity. Hence, a comparison between the Boltzmann and Landauer approaches
still remains an open and an important issue.

The main purpose of the present study is to compare the exact Landauer
tight-binding (TB) conductivities with those given by the standard Boltzmann
approach within the Born approximation as well as with those given by Eq. (%
\ref{Stauber}). As in the standard Boltzmann approach the density dependence is different
for long- and short-range scatterers, one of our aims is to investigate whether the\ TB
calculations also give different density dependencies for these scattering mechanisms.
Finally, the Born approximation is valid for the case of weak scattering when the wave
functions remains unperturbed. It is not however apparent the condition of a weak
scattering is satisfied in a parameter range typical for realistic systems. By comparing
the exact\ TB calculations with those based on the predictions of the standard Boltzmann
approach we test the applicability of the later to realistic systems.

\section{Theory}

We calculate conductivity of the graphene sheets using the standard $p$%
-orbital nearest-neighbor tight-binding Hamiltonian for non-interacting
electrons for zero temperature, $H=\sum_{i}V_{i}\left\vert i\right\rangle
\left\langle i\right\vert -t\sum_{i,j}\left\vert i\right\rangle \left\langle
j\right\vert $, with the hopping integral $t=2.7$ eV.\cite%
{Castro_Neto_review} We utilize a model for screened scattering centers of
the Gaussian shape (Eq. (\ref{Gauss_pot})) commonly used in the literature%
\cite{Bardarson,Lewenkopf,ABDS,Klos}, The correlator of the potential (\ref%
{Gauss_pot}) has the form, \cite{Bardarson,Lewenkopf,ABDS} $\left\langle
V_{i}V_{j}\right\rangle =\frac{K(\hbar v_{F})^{2}}{2\pi \xi ^{2}}\exp \left(
-\frac{|\mathbf{r}{_{i}}-\mathbf{r}_{j}|^{2}}{2\xi ^{2}}\right) ,$ where the
dimensionless impurity strength is described by the parameter $K\approx
40.5n_{imp}(\delta /t)^{2}(\xi /\sqrt{3}a)^{4}$ given by the screening
length $\xi $, the potential strength $\delta $, and the relative impurity
concentration $n_{imp}$ ($a$\textbf{\ }being the carbon-carbon distance, and
$v_{F}$ is the Fermi velocity). For realistic graphene samples $%
K=1\sim 10.$ \cite{Bardarson,Lewenkopf,ABDS,Klos}. In our calculations we use the
screening length $1\leq \xi /a\leq 8,$ spanning the range between the short-range
potential ($\xi =a)$, and the long-range potential ($\xi =8a)$ that varies smoothly on
the scale of a lattice constant. Note that we also performed calculations for the $\delta
$-scattering, and the results obtained are, as expected, practically identical to those
obtained for the case of the Gaussian disorder with $\xi =a.$

The conductance $G$ and the electron density $n$ are computed with the aid
of the recursive Green's function technique \cite{Xu,Xu_bilayer,Klos}. We
assume that the semi-infinite leads are perfect graphene strips of the width
$W$, and the device region is a rectangular graphene strip $W\times L$ where
the impurity potential is defined. The zero-temperature conductance $G$ is
given by the Landauer formula $G=\frac{2e^{2}}{h}T,$ where $T$ is the total
transmission coefficient between the leads. Then we calculated the
conductivity $\sigma =\frac{L}{W}G,$ the electron density $%
n=\int_{0}^{E_{F}}dE\;D(E),$ the mobility $\mu =\frac{\sigma }{en},$ and the
mean free path (mfp) $\mathrm{mfp}=\frac{h}{2e^{2}}\frac{\sigma }{\sqrt{\pi n%
}}$ as a functions of the Fermi energy $E_{F}$. The density of states (DOS) $%
D(E)$ is computed by averaging the local density of states (LDOS) over the
whole device area. The LDOS is given by the diagonal elements of the total
Green's function.\cite{Xu}. Because of computational limitations we study
the strips with $L/W>1$ (In most calculations we use $L=368$ nm (3000 sites)
and $L/W\approx 6).$ We however checked that the obtained results are
insensitive to the aspect ration $L/W$ as soon as $L/W>1.$

The results of the tight-binding calculations are compared to the
predictions obtained within the standard Boltzmann approach within the Born
approximation for the scattering potential (\ref{Gauss_pot}). The
conductivity of graphene sheet is given by\cite%
{Nomura,Ando2006,Hwang2007,Stauber,Katsnelson,Vasko} $\sigma =e^{2}\tau
D(E_{F})\frac{v_{F}^{2}}{2}$, where the DOS $D(E_{F})=2E_{F}/\left( \pi \hbar
^{2}v_{F}^{2}\right) ,$ the dispersion relation $E=$ $\hbar v_{F}k,$ and $k=\sqrt{\pi
n}$, with $v_F=\frac{3at}{2\hbar}$ being the Fermi velocity. The scattering rate within
the Born approximation
reads $\tau ^{-1}\mathbf{=}\frac{2\pi }{\hbar }D(E)\frac{1}{4}\int_{0}^{\pi }%
\frac{d\theta }{\pi }\frac{1-\cos ^{2}\theta }{2}|U_{q}|^{2},$ where $U_{q}$
is the Fourier transform of the scattering potential, $q=|\mathbf{k}-\mathbf{%
k}^{\prime }|=2k\sin \frac{\theta }{2},$ where $\theta $ is the angle
between the initial and final states $\mathbf{k}$ and $\mathbf{k}^{\prime }.$
Using the Wiener-Kitchine theorem for the correlator, we obtain the Fourier
transform of the Gaussian potential, $|U_{q}|^{2}=K(\hbar v_{F})^{2}\exp
\left( -q^{2}\xi ^{2}/4\right) ,$ which leads to the expression for the
conductivity\cite{ABDS,Vasko},
\begin{equation}
\sigma =\frac{4e^{2}}{h}\frac{\pi n\xi ^{2}e^{\pi n\xi ^{2}}}{KI_{1}(\pi
n\xi ^{2})}\sim
\begin{array}{c}
\text{const, }\pi n\xi ^{2}\ll 1,\text{ } \\
n^{3/2},\;\pi n\xi ^{2}\gg 1%
\end{array}%
,  \label{Boltzmann conductivity}
\end{equation}%
where $I_{1}$ is the modified Bessel function.

\begin{figure}[tbp]
\includegraphics[width=\columnwidth]{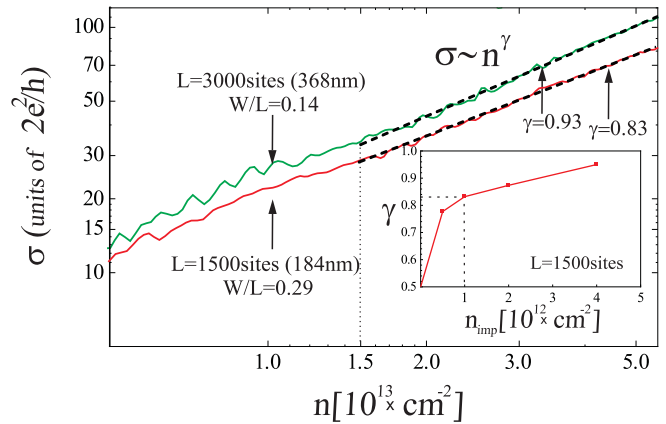} 
\caption{(a) The conductivity vs the electron density for two representative
graphene strips with different lengths $L=3000$ and $1500$ sites for the
case of the short-range scattering ($\protect\xi =a$) with $n_{imp}=10^{12}$
cm$^{-2} $. Each curve is averaged over 8 impurity configurations. The
impurity strength $\protect\delta =0.86t$. (b) The inset shows the
dependence $\protect\gamma =\protect\gamma (n)$.}
\label{fig:ballistic-diffusive}
\end{figure}

\section{Results and discussions}

The Boltzmann predictions for the density dependence of the conductivity are
valid in the diffusive transport regime when the mean free path is larger
than a system size. Let us therefore first discuss a transition from the
ballistic to diffusive regime focussing on the short-range scattering, $\xi
=a$. In a purely ballistic regime (no impurity scattering) the conductivity
follows the density dependence $\sigma \sim n^{\gamma }$ with $\gamma =%
{\frac12}%
$. \cite{Klos} It has been demonstrated for the case of the long-range
Gaussian scatterers that with the increase of the system size the exponent $%
\gamma $ gradually increases from its ballistic value reaching the value $%
\gamma =1$ in the diffusive regime.\cite{Klos} Figure \ref%
{fig:ballistic-diffusive} shows the dependence $\sigma =\sigma (n)$ for the case of the
short-range scattering calculated within the TB approach. The conductivity shows the same
behavior as for the case of the long-range scattering\cite{Klos} with $\gamma $
increasing from
$\frac12$
in the ballistic regime to $\gamma =1$ in the diffusive regime as the size
of the system or the impurity concentration increases. This obtained density
dependence in the diffusive regime ($\gamma =1)$ is in a stark contrast with
the standard Boltzmann predictions for the $\delta $-impurity scattering,
Eq. (\ref{short}), when $\sigma $ is expected to be density independent ($%
\gamma =0)$.

\begin{figure}[tbp]
\includegraphics[width=\columnwidth]{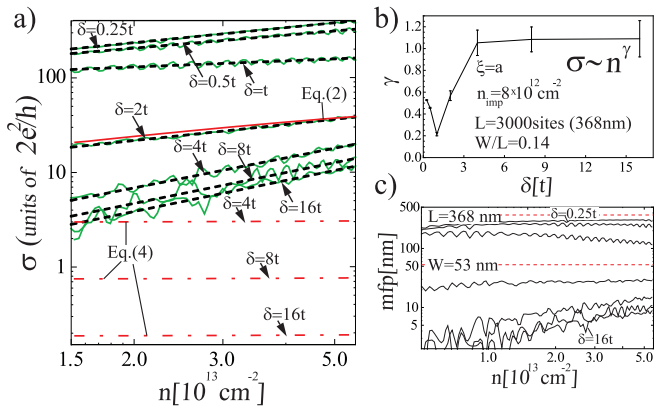} 
\caption{The conductivity of the graphene sheets vs electron density for the
short-range Gaussian potential with $\protect\xi =a$ for different potential
strength $\protect\delta $. Each curve is averaged over 8 impurity
configurations. Straight dashed lines show fitting $\protect\sigma \sim n^{%
\protect\gamma }.$ Red solid and dot-dashed lines show predictions based respectively on
Eqs. (\protect \ref{Boltzmann conductivity}) and (\protect\ref{Stauber}). (b) The
dependence $\protect\gamma =\protect\gamma (\protect\delta ).$ (c) The m.f.p. vs the
electron density $n.$ } \label{fig:short-range}
\end{figure}

A more detailed comparison between the TB and Boltzmann calculations for the short-range
scatterers for different impurity strengths $\delta $ is presented in Fig.
\ref{fig:short-range} (a)-(b). As expected, for very weak scattering ($\delta
\lesssim 0.5t$) the transport is in the ballistic regime with $%
\gamma \approx 1/2.$ This is fully consistent with the calculated m.f.p.
which is comparable to the largest dimension of the system $L$, see Fig. \ref%
{fig:short-range} (c). For the case of strong scattering ($\delta \gtrsim 2.5t$) the
system is in the diffusive transport regime when the calculated m.f.p. is smaller than
the smallest dimension of the system $W$. In this regime the exponent $\gamma $ saturates
to 1, see Fig. \ref{fig:short-range} (b). Figure \ref{fig:short-range} (a) also shows the
conductivity calculated on the basis of the standard Boltzmann approach, Eq.
(\ref{Boltzmann conductivity}), as well as given by Eq. (\ref{Stauber}). The Boltzmann
theory predicts that $\gamma =0$ which is in qualitative disagreement with the
numerically calculated exponent $\gamma \approx 1.$ Boltzmann predictions are also
quantitatively different from the tight-binding calculations with $\sigma ^{Boltz}\ll
\sigma ^{TB}$ (note the logarithmic scale of the figure). At the same time, we find that
the TB calculations are in a good qualitative and even reasonable good quantitative
agreement with Eq. (\ref{Stauber}) predicting quasilinear density dependence of the
conductivity. Why does the standard Boltzmann approach fail to describe the conductivity
of the system at hand? Following Stauber\textit{\ at al}.\cite{Stauber}, we believe this
is because the scattering rate $\tau $ in the standard Boltzmann approach is calculated
in the Born approximation, with unperturbed clean-graphene wave functions. Apparently,
this approximation is applicable in the case of weak perturbations, but cannot be applied
for strong scattering potential. In contrast, the approach proposed by Stauber\textit{\
at al}. uses wave functions for a hard-wall barrier, appropriate in the case of strong
scattering.

Let us now discuss the transition regime between the ballistic and diffusive behavior
when $W<$ m.f.p. $<L$ (which corresponds to the impurity strength in the vicinity of
$\delta \approx t)$ where the exponent $\gamma $ shows a pronounced minimum dropping to
$\gamma \approx 0.2$ for $\delta \approx t$ . We are not aware of any theories addressing
this transition regime corresponding to the quasiballistic transport. We speculate,
however, that this peculiar behavior, with the conductivity becoming weakly dependent on
the concentration, might be related to the corresponding Boltzmann prediction of $\gamma
=0$ in the diffusive regime (even though the Boltzmann theory is
not formally applicable in the case under consideration, with $W<$ mfp $%
<L).$

\begin{figure}[tbp]
\includegraphics[width=\columnwidth]{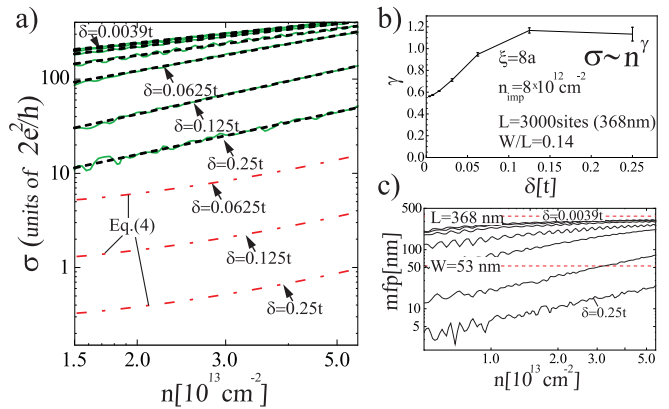} 
\caption{(a) The conductivity of the graphene sheets vs electron density for
the long-range Gaussian potential with $\protect\xi =8a$ for different
potential strength $\protect\delta $. Each curve is averaged over 8 impurity
configurations. Straight lines show fitting $\protect\sigma \sim n^{\protect%
\gamma }.$ Dashed lines show predictions based on Eq. (\protect\ref%
{Boltzmann conductivity}). (b) The dependence $\protect\gamma =\protect%
\gamma (\protect\delta ).$ (c) The m.f.p. vs the electron density $n.$ }
\label{fig:long-range}
\end{figure}

Let us finally compare the tight-binding and Boltzmann calculations for the
long-range Gaussian scatterers with $\xi =8a$ for different impurity
strengths, see Fig. \ref{fig:long-range}. As for the case of the short-range
scatterers the TB calculations exhibit the ballistic behavior (m.f.p. $%
\approx L$) with $\gamma =1/2$ for weak scattering and the diffusive behaviour (m.f.p.
$\lesssim W$) with $\gamma \approx 1$ for the strong scattering. Again, the result
obtained in the diffusive regime, $\gamma \approx 1$, is qualitatively different from the
corresponding Boltzmann prediction. Indeed, for the latter case the exponent $\gamma $ is
poorly defined because for the considered
density interval $\pi n\xi ^{2}\approx 1$ which, according to Eq. (\ref%
{Boltzmann conductivity}), corresponds to the transition regime between two asymptotes
$\sigma =$const and $\sigma \sim n^{3/2}.$ Besides, the Boltzmann and the tight-binding
calculations disagree even quantitatively
with $\sigma ^{Boltz}\ll \sigma ^{TB}$ (Note the logarithmic scale of Fig. \ref%
{fig:long-range} (a)).

The opposite limit $\pi n\xi ^{2}\gg 1$ (when $\sigma \sim n^{3/2}$) was considered by
Adam \textit{et al.} \cite{ABDS} who found a good qualitative and quantitative agreement
between the Landauer-type and the Boltzmann calculations. This regime (in contrast to the
regime $\pi n\xi ^{2}\lesssim 1$ considered here) corresponds to high electron energies
and smooth potential (with large $\xi ) $ when the scattering is weak and the Boltzmann
theory within the Born approximation is therefore justified. However, the density
dependence predicted by Eq. (\ref{Boltzmann conductivity}), $\sigma \sim n^{3/2},$ has
never been observed in any experiment. In contrast, the dependence $\sigma \sim n$,
obtained in the TB calculations for the regime $\pi n\xi ^{2}\lesssim 1$ considered here,
is in agreement with the majority of experimental findings. We regard this as a strong
indication that the regime appropriate for realistic graphene samples is $\pi n\xi
^{2}\ll 1$. As demonstrated here, in this parameter range the results obtained in the
standard Born approximation disagree both quantitatively and qualitatively with those
obtained by the exact\ TB calculations. This therefore questions the validity of the
standard Boltzmann predictions within the Born approximation for realistic graphene
sheets.

\section{Conclusions}

(\textit{i}) In relatively small systems the transport is in the ballistic regime for
both short- and
long-range scatterers, and the density dependence of the conductivity is $%
\sigma \sim n^{%
{\frac12}%
}$. As the system size (or the impurity concentration) increases the ballistic regime
becomes superseded by the diffusive regime, in which the\ TB calculations predict the
same linear density dependence, $\sigma \sim n,$ for both short- and long-range
scattering.

(\textit{ii}) In the case of short-range potential the obtained linear dependence is in
quantitative and qualitative disagreement
with the standard Boltzmann predictions within the Born approximation, Eqs. (%
\ref{short}) and (\ref{Boltzmann conductivity}), but in agreement
with the predictions going beyond the Born approximation, (\ref{Stauber}%
).

(\textit{iii}) Even for the long-range Gaussian potential the standard Boltzmann
predictions (\ref{Boltzmann conductivity}) are in quantitative and qualitative
disagreement with the TB results in the parameter range corresponding to realistic
systems ($\pi n\xi ^{2}\lesssim 1$ regime).  This questions the applicability of the
predictions based on the standard Boltzmann theory for conductivity in graphene which are
widely used for interpretation of experimental data.

Discussions with N. M. R Peres, T. Heinzel, H. Xu, A. Shylau, F. Vasko are
greatly appreciated. I.V.Z. acknowledge support from the Swedish Research
Council (VR).


\end{document}